\documentclass[reprint,amsmath,amssymb,aps,prl]{revtex4-2}

\usepackage{graphicx}
\usepackage{dcolumn}
\usepackage{bm}
\usepackage[utf8]{inputenc}
\usepackage{hyperref}
\usepackage{float}
\usepackage[abs]{overpic}
\usepackage{gensymb}
\usepackage{upgreek}
\usepackage{color}
\usepackage{xcolor}
\usepackage{comment}
\usepackage{mathtools}

\makeatletter
\renewcommand\@makecaption[2]{%
	\par
	\vskip\abovecaptionskip
	\begingroup
	\small\rmfamily
	\begingroup
	\samepage
	\flushing
	\let\footnote\@footnotemark@gobble
	\@make@capt@title{#1}{#2}\par
	\endgroup
	\endgroup
	\vskip\belowcaptionskip
}

\begin{document}
	
\preprint{APS/123-QED}

\title{Discrimination of Chiral and Helical Contributions to Raman Scattering of Liquid Crystals using Vortex Beams}

\author{Silvia Müllner$^{1}$}
\author{Florian Büscher$^{1}$}
\author{Angela Möller$^{2}$}
\author{Peter Lemmens$^{1}$}

\affiliation{$^{1}$Institute for Condensed Matter Physics, University of Technology Braunschweig, D-38106 Braunschweig, Germany\\   
	$^{2}$Department of Chemistry, JGU Mainz, D-55128 Mainz, Germany }

\date{\today}

\begin{abstract}We use vortex photon fields with orbital and spin angular momentum to probe chiral fluctuations within liquid crystals. In the regime of iridescence with a well-defined pitch length of chirality, we find low energy Raman scattering that can be decomposed into helical and chiral components depending on the scattering vector and the topological charge of the incident photon field. Based on the observation of an anomalous dispersion we attribute quasi-elastic scattering to a transfer of angular momenta to roton-like quasiparticles. The latter are due to a competition of short-range repulsive and long-range dipolar interactions. 
Our approach using a transfer of orbital angular momentum opens up an avenue for the advanced characterization of chiral and optically active devices and materials. 




\end{abstract}

\maketitle

The competition of long-range dipolar interactions with short-range repulsion is known to lead to interesting phase diagrams with a variety of exotic phases and fluctuations. Examples are found in supra-fluid helium \cite{NeutronscatterinHe}, trapped dipolar Bose-Einstein condensates (BEC)\cite{dipolargas, Rotonsoftening}, exciton states of 2D semiconductors, and micro-mechanical networks with further distance force elements \cite{metamaterial}. Also phonons of chiral semi-metals and layered micro-polar compounds are discussed in this context \cite{Liu-21,Kishine-20,Baghda-22}. Here, we highlight inelastic light scattering using Laguerre-Gaussian (LG) laser beams with helical phase front of the photon field. Using such structured light it seems feasible to gain novel insight into chiral fluctuations. We propose chiral liquid crystals (LC) to realize a model and table top system for the study of exotic phases. LC have large and anisotropic optical polarizabilities that are advantageous for optical probes. The observed smectic and chiral phases that can be tuned by temperature in a convenient regime, offer pronounced fluctuations that origin from the interplay of short-range repulsion with long-range dipolar interactions.
	
In this report we analyze the phase front of scattered Raman photons and discriminate the transfer of orbital angular momentum (OAM) versus spin angular momentum (SAM) to the chiral phase of LC. Interestingly, we observe two different signals with different line-width and energy. For a small momentum transfer there exist finite-energy Lorentzian fluctuations with a dominantly helical polarization (SAM). With larger momentum transfer, we observe low-energy Gaussian fluctuations with chiral polarization (OAM). This constitutes a previously unobserved dichotomy in Raman scattering (RS). We attribute the anomalous low energy modes to chiral, Roton-like quasi particles, in analogy to supra-fluid Helium. 

LG modes with OAM and SAM components of the vortex beam are often referred to as structured or twisted light. This is due to their helical phase front along the direction of propagation. OAM photon fields are beneficial for quantum information transfer, optical tweezers, super-resolution microscopy and enhanced sensing of molecular optical activity \cite{Rubinsztein, tweezers}. The discussion of light matter interaction and OAM-based optical activity remains controversial since positive \cite{enhanced1,enhanced2,enhanced3} as well as negative results \cite{Loeffler1} exist. However, it is confirmed and widely accepted that the phase of LG modes does not couple to dipole matrix elements in the paraxial regime \cite{Araoka,Wozniak}. More recently, higher order effects, resonances, and focussing-induced spin-orbit coupling are discussed\cite{Forbes-21b, spin-orbit}. 

LG laser beams are paraxial solutions of a wave equation and consist of both spin (helicity, SAM) and orbital (chirality, OAM) components. We have prepared such chiral/helical photon fields with SAM and OAM of different topological charge $\ell=\pm1,\pm2,\pm3,\pm4$ by transmission of left-handed (LH) and right-handed (RH) circular polarized light (CPL) through corresponding vortex phase retarders (Thorlabs). Despite a normalizing factor, the operator mode expansion for the transverse electric displacement field for a LG beam is given by \cite{Forbes-21}: 
\begin{equation}
	\begin{split}
		\label{eq:2}
		d_i^{ \perp }=i\sum_{k, \sigma, \ell, p} \bigg[& \hat{e}^\sigma(k\hat{z})  f_{|\ell|, p}(r){a_{|\ell|,p}}^\sigma (k\hat{z})e^{i(kz+\ell\phi)}\\ & -H.c.\bigg].
	\end{split} 
\end{equation}
H.c. is the Hermitian conjugate, $\hat{e}^\sigma$ the electric polarization unit vector, ${a_{|\ell|,p}}^\sigma (k\hat{z})$ the annihilation operator, and $f_{|\ell|,p}(r)$ the radial distribution function. $e^{i(kz+\ell \phi)}$ is the azimuthal dependent phase factor accountable for the OAM with the topological charge $\ell$.

We use the well established, left-handed LC cholesteryl nonanoate, (C$_{36}$H$_{62}$O$_{2}$) (Sigma-Aldrich), as a model system. Its chiral nematic phase can be rationalized as a chiral, photonic crystal with anomalous real and imaginary part of the dielectric constant. We thereby achieve a localization of light, similar to a metamaterial. The building block of this functionality is the highly polarizable, anisotropic molecule that tends to spiral due to the competition of shape, van der Waals, and dipole-dipole interactions. Previous studies have shown that LC have exceptional, non-linear optical properties\cite{Khoo-09} allowing the realization of numerous photonic devices \cite{Zhang-21}, including lasers with topological effects\cite{Papic-21}. Therefore, they are also candidate materials for a search of OAM light matter coupling. 

Our sample is placed in a rectangular, optical cuvette (10mm, Hellma Analytics) mounted on a thermo-electric Peltier cooler/heater (Belektronig BTC-LAB-A2000). The optical setup allows experiments in back scattering and transmission geometry with the same sampling optics (aperture). The latter defines the scattering vectors of the experiments. 

Previous circular dichroism experiments on LC have shown that a possible effect of OAM must be very small \cite{Araoka,Loeffler1}. Therefore, we probe the ratio of RS with opposite helicities ({$I_{LH}/I_{RH}$) or opposite chiralities ($\pm \ell$) ({$I_{\ell=1}/I_{\ell=-1}$). This approach is known from Raman optical activity (ROA). It allows to discriminate effects due to a transfer of SAM or OAM, respectively \cite{Forbes-19}. Contributions independent of angular momentum cancel out. We refer to Ref. \cite{Florian} for more general RS experiments. All back scattering RS experiments use a focused laser beam with $\lambda_{\text{exc}}=532.1$ nm excitation wavelength, $P_{\text{exc}}=4-15$ mW and an angle of incidence of 20$^\circ$ with respect to the normal of the sample cuvette/liquid interface. This reduces elastic scattering. In transmission RS we use a corresponding angle of $7^\circ$.  

 \begin{figure}
	\includegraphics[width=65mm]{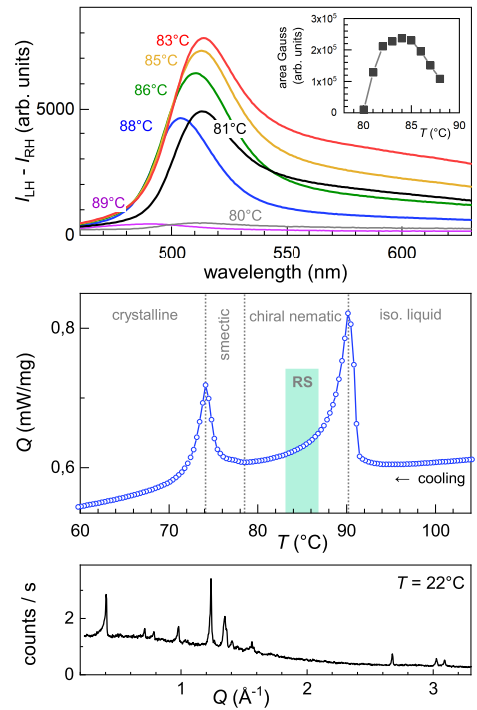}
	\caption{\label{fig:Fig1}(a) Spectroscopically resolved reflectivity $I_{\text{LH}}-I_{\text{RH}}$ for different temperatures. The inset (squares) shows the intensity of a maximum fitted by a Gaussian. (b) Differential scanning calorimetry (DSC). The box corresponds to the regime of our OAM RS experiments. (c)X-ray diffraction graph taken at room temperature after multiple thermal cycling.}
\end{figure}

We determine the temperature and frequency interval of iridescence in the chiral nematic phase using spectroscopically resolved reflectivity measurements (Hamamatsu, mini-spectrometer). The intensity difference  $I_{\text{LH}}-I_{\text{RH}}$, using a RH and LH CPL in Fig.\ref{fig:Fig1}(a) shows a sharp intensity maximum in the temperature range between \mbox{81 $^\circ$C and 88 $^\circ$C} (see inset) with a line-width of 30 nm FWHM. The maximum energy shifts towards shorter wavelength with increasing temperature. This reflects the temperature dependent pitch length, {\it l(T)} of the chiral liquid. Furthermore, we find that the chiral LC is predominantly a LH reflector and RH giving only a background signal. This maximum in the regime of iridescence is a significant fingerprint of resonant light matter interactions. We find that the peak position of the maximum wavelength also dependents on the incident angle of light with respect to sample normal. With larger angles the maximum decreases in wavelength. Light diffracted into the liquid probes the respective projection of the pitch length with respect to the {\it k}-vector of light. This implies the need of well defined aperture of the sampling optics as it defines the range of scattering vectors and pitch lengths that is integrated on.

Fig. \ref{fig:Fig1}(b) depicts differential scanning calorimetry (DSC-3, Mettler-Toledo) to determine the thermodynamic phase diagram using a cooling rate of 20 K/min. A box in the figure shows the temperature regime of iridescence and our RS experiments, described below. Thereby, we ensure that the observed effects are not due to the phase transitions. Room temperature powder x-ray diffraction (STADI-P, STOE\&Cie) data were collected after multiple thermal cycling to verify the phase stability of the samples (\mbox{Fig. \ref{fig:Fig1} (c)}).

In Fig.~\ref{fig:Fig2}(a) divided RS intensities, $I_{\ell=2}/I_{\ell=-2}$, are shown as a function of temperature. They are consecutively shifted for convenience. We observe no effect for large Raman shifts and the spectral division results to unity, i.e. spectra of different OAM are identical. At low frequencies ($<150$ cm$^{-1}$), however, a Gaussian line-width scattering surplus emerges within the temperature interval $T_{\text{res}}\approx 83^{\circ}$C$-87^{\circ}$C. The quite small temperature window of observation points to an underlying resonance effect. The later is very probably related to the maximum in reflectivity, \mbox{(see Fig. \ref{fig:Fig1}(a))}, that exists as function of temperature and wavelength due to the T-dependence of {\it l(T)}. We observe no vibrational modes in $I_{\ell=2}/I_{\ell=-2}$. This is rationalized by a negligible spin-orbit interaction in these light element systems. 

There remain two important questions: (a) What is the role of the total angular momentum of light, i.e. can we differentiate the helicity vs. chirality of the chiral phase of LC; (b) Is there a dependence on the scattering vector of the experiments. In other words, is there a dispersion of the observed RS that could be interpreted as a quasi-particle. The first question is particularly relevant as structured light ($\ell\ne0$) contains helicity (SAM) as well as chirality (OAM) of the involved photons. In the generation of a finite OAM, circular polarized light (LH or RH) is transmitted through a wave plate, leading to a reversal of handedness together with the generation of OAM of the photon field.

In Fig. \ref{fig:Fig2}(b) we compare data with SAM (RH/LH) only and SAM combined with OAM of varying magnitude ($\ell=\pm 1,\pm 2,\pm 3$,$\pm 4$) at the temperature of highest selective reflectivity, $T_{\text{s}}= 84^\circ$C. All polarizations show a low energy surplus scattering (QES) with a 150 cm$^{-1}$ line-width. Nevertheless, its intensity is largest for $\ell=+2/-2$ and smallest for $\ell=+4/-4$, see \mbox{Fig. \ref{fig:Fig2}(c)}. The intensity of purely helical polarization (LH/RH, $\ell=0$) is a factor $0.7$ smaller than $\ell=+2/-2$. We notice that for $\ell=+2/-2$ the phase factor $e^{i \ell \phi}$ of the Laguerre-Gaussian (see Eq.\ref{eq:2}) has an additional resonance with the pitch length of the chiral liquid, leading to the maximum intensity.

\begin{figure}
	\begin{ruledtabular}
		\includegraphics[width=65mm]{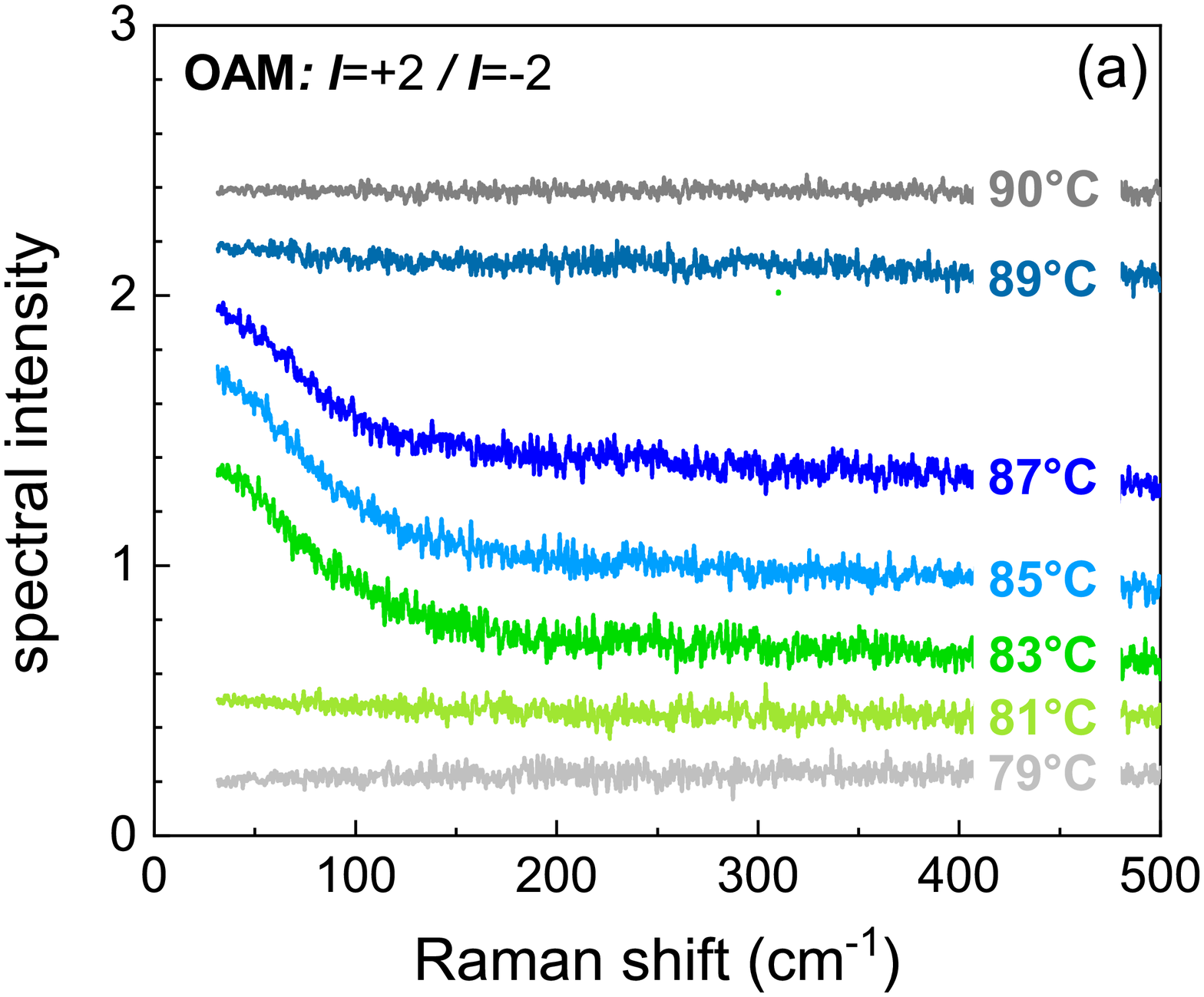}
		\includegraphics[width=65mm]{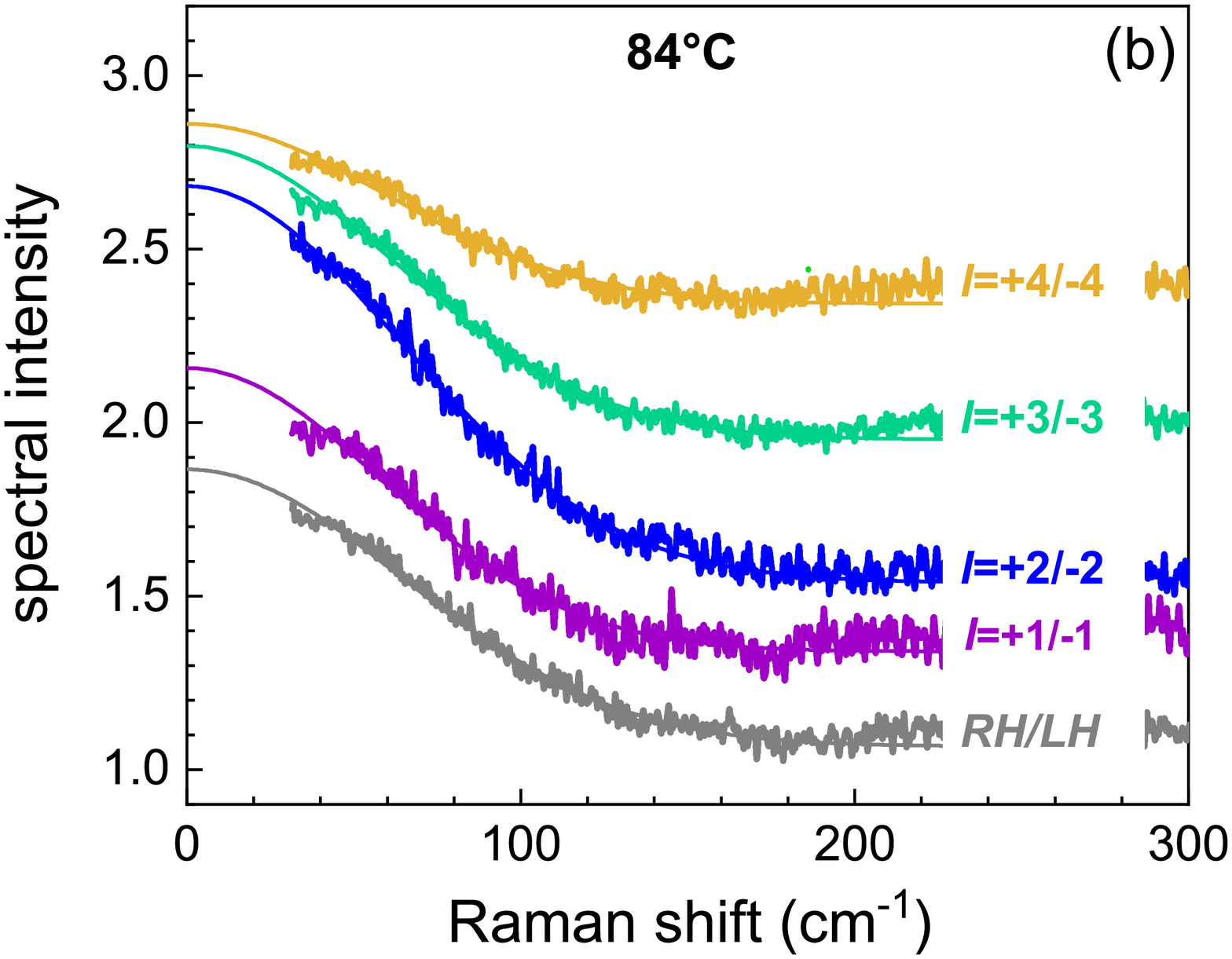}
		\includegraphics[width=65mm]{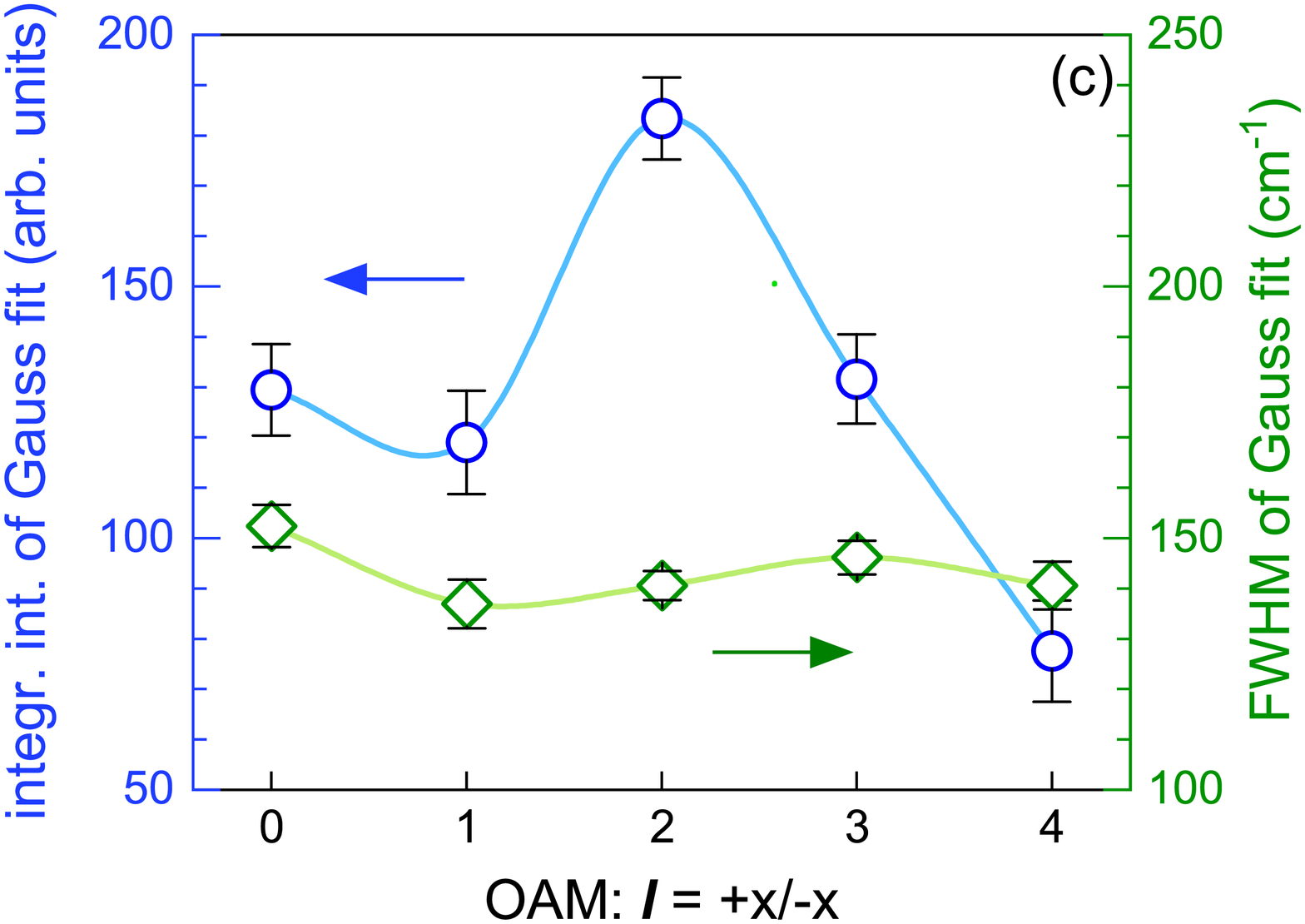}	
		\caption{\label{fig:Fig2}(a) Divided OAM Raman spectra with $\ell=+2/-2$ as a function of temperature, shifted for convenience. \label{fig:Fig3b}(b) Divided OAM Raman spectra comparing helical polarization (RH/LH) only and different OAM states at $T_s$=84$^{\circ}$C. Full lines show fits using a Gaussian centered at $E\approx0$cm$^{-1}$. (c) Integrated Gaussian intensity (blue circles) and FWHM line-width (green diamonds) of (b). OAM = 0 corresponds to helical only, RH/LH, polarizations.}
	\end{ruledtabular}
\end{figure} 

\begin{figure}
	\includegraphics[width=65mm]{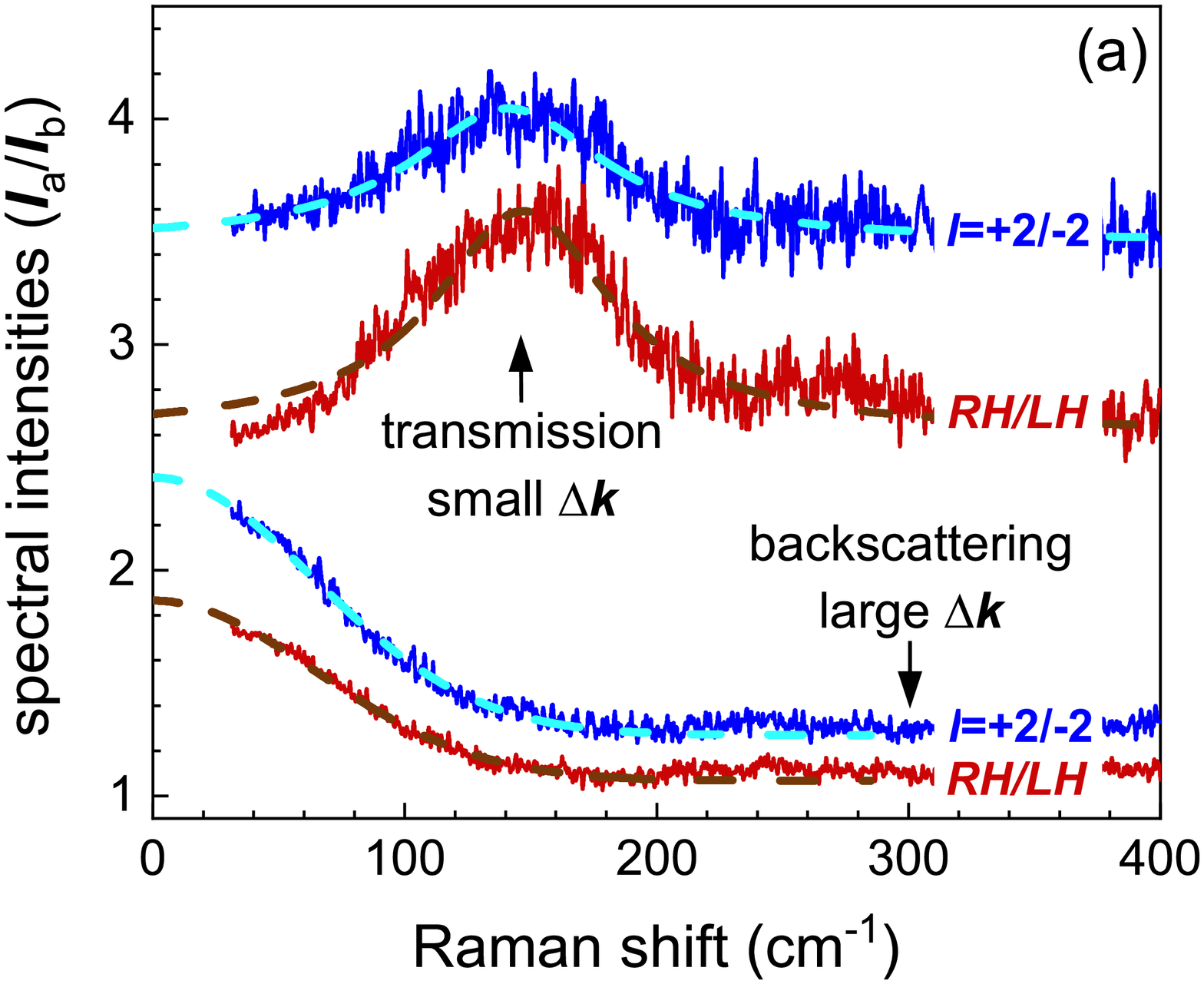}
	\includegraphics[width=65mm]{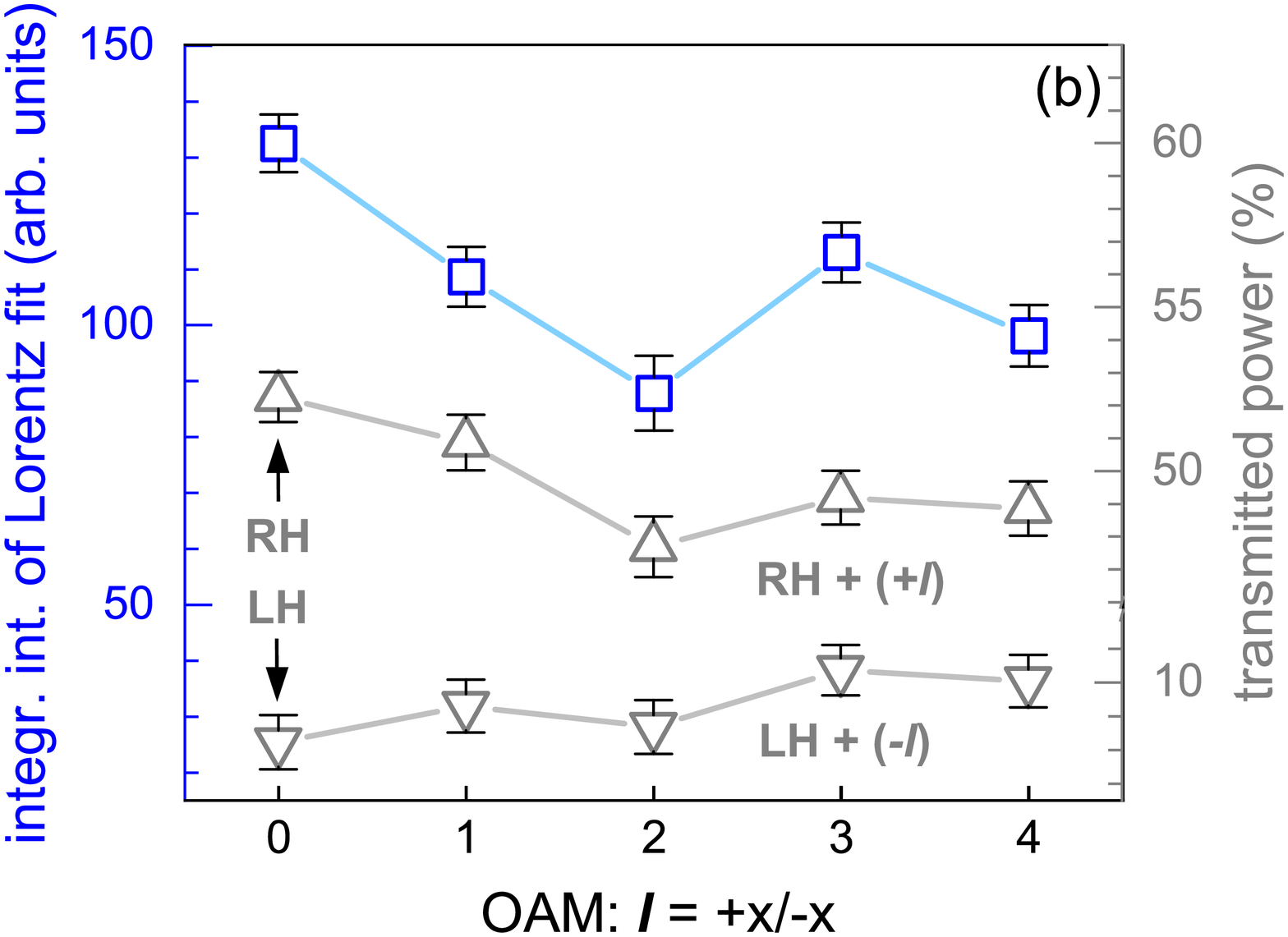}
	\includegraphics[width=55mm]{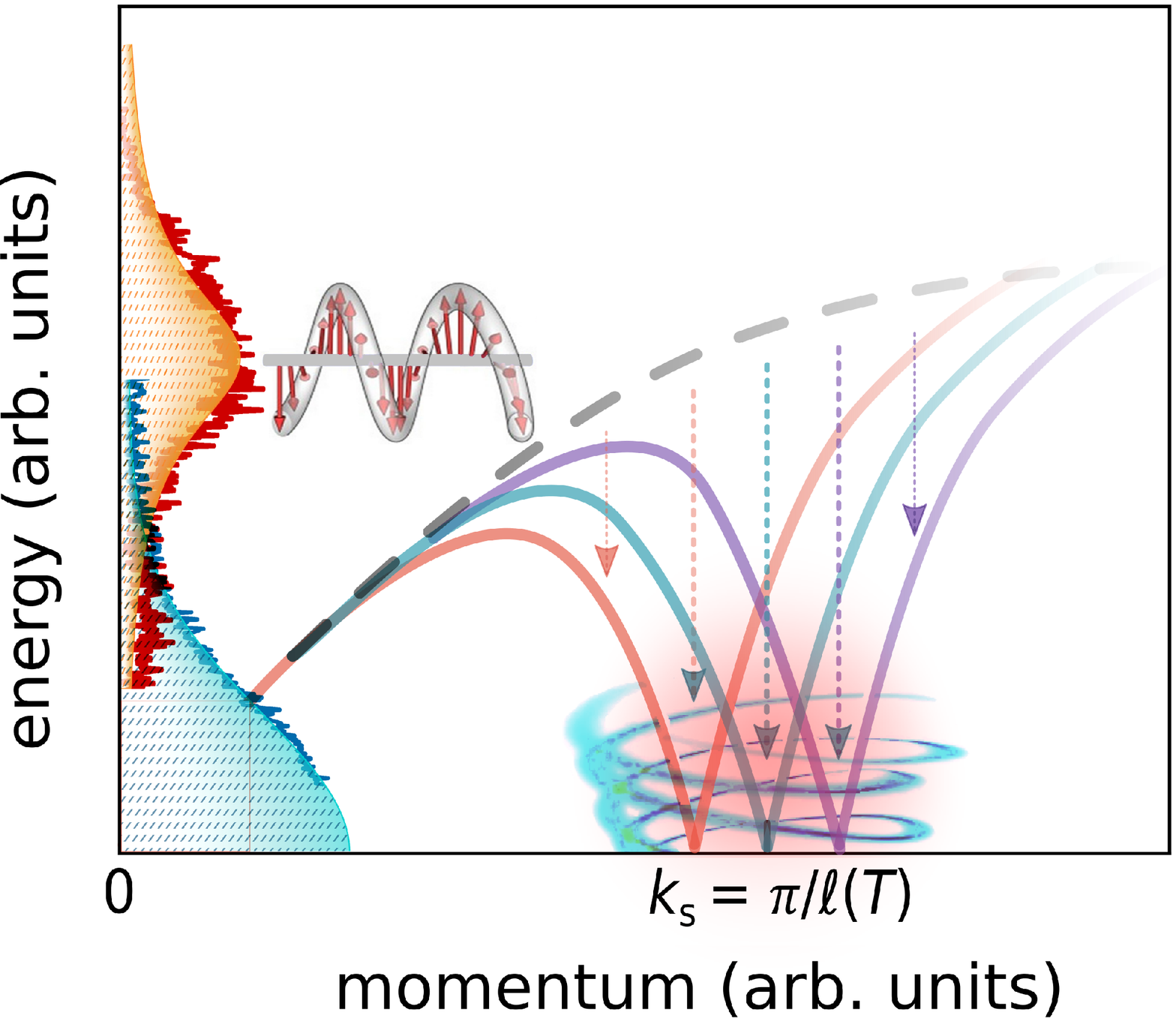}
	\caption{\label{fig:Fig3}(a) Comparison of transmission RS (upper curves), back-scattering RS (lower curves) with two polarizations, and fits to finite-energy Lorentzian and quasi-elastic, Gaussian functions, respectively. (b) Transmitted power measurements as function of light polarization at $T_{s}$=84$^{\circ}$C. (c) Sketch of an acoustic dispersion branch, T-dependent roton minima, and suggested RS signals.}
\end{figure}

In the following we compare transmission RS data with the previous back-scattering data searching for possible dispersion effects. Transmission RS has a vanishing scattering vector, {$\Delta k \approx 0$}, while back-scattering involves potentially large scattering vectors, $\Delta k = k_i-k_s \approx 2 k_i$, due to the reversed $k$-vector. Furthermore, we consider that in the regime of iridescence the wavelength of incident photons is identical with \textit{l(T=T${s}$)} of chirality, the latter giving a chiral Brillouin zone of $\Delta k=2\pi/l(T_{s})$. Also in this sense the scattering vector of the back-scattering experiment is large. 

In Fig. \ref{fig:Fig3} (a) we find that transmission data show a finite energy maximum, \mbox{$E_{\text{center}}\approx150$ cm$^{-1}$}. With a line-shape close to a Lorentzian, this signal resembles to a broadened phonon or phonon density of states. This phenomenology is actually completely different from the previous back-scattering data showing a broad, Gaussian distribution of very small energies. In Fig. \ref{fig:Fig3} (b) we study the integrated intensity of the above maxima (blue open squares) as function of SAM and OAM. It shows a maximum for $\ell$=0 and a minimum for $\ell=+2/-2$. Thereby, it is opposite to the one of back-scattering data, Fig. \ref{fig:Fig2}(c). Finally, Fig. \ref{fig:Fig3} (b) (open triangles) shows the transmitted power at $\lambda$=532~nm at $T=T_{s}$. We observe a larger transmission for all RH compared to LH helicities. This is a nice and independent proof of iridescence being based on helicity and its transformation as an axial vector. Therefore, replacing back-scattering by transmission reverses the effect on the measured intensities.

To rationalize these data, we will now discuss qualitatively the acoustic dispersion of a system with long-range dipolar interactions, see Fig. \ref{fig:Fig3} (c). Switching on the interactions, the conventional linear branch at small momenta crosses over to a maximum and softens to a minimum. Excitations close by are called \textit{rotons} and the systems reaches a roton instability if the minimum touches $E=0$. The term \textit{roton} has shown a remarkable development from a nonlocal measure of vorticity, an incipient crystallization \cite{Noziere-04,Noziere-09}, to a chirality-induced cross correlation of lattice to electronic degrees of freedom\cite{Kishine-20}. The roton minimum can be understood as a fingerprint of complexity related to the non-local pair potential and induced non-integrability.

For a chiral liquid we expect a minimum at a characteristic \textit{k$_{s}$(T})} = $\pi/\textit{l(T)}$. \textit{k$_{s}$} will shift with increasing temperatures to larger momenta, as derived from Fig.\ref{fig:Fig1}(a)). Such a dependence also implies that thermal fluctuations are correlated with helical fluctuations. This constitutes a novel contribution to light matter coupling via the strongly nonlinear electronic polarizability of their molecular components. As mentioned before, \textit{k$_{s}$} also defines the boarder of the chiral Brillouin zone and it is locked to the RS scattering vector $\Delta k$. Therefore, we expect that iridescence leads to a down-folding of the quasi-particle dispersions to the Brillouin zone center, an important aspect for optical experiments. 

In Fig. \ref{fig:Fig3} (c) we sketch the resulting Raman signals at $k\approx 0$, i.e. shaded Lorentzian and Gaussians located at different energies. A broader, finite energy maximum will result from multi-particle RS due to the large density of states at the dispersion maxima. This RS follows from the fluctuation-dissipation theorem and is based on a conventional density-density correlation function. Therefore, we expect as a linewith a Lorentzian or related superpositions. Presently the exact location of this signal with respect to the dispersion maximum is unknown and could be around 1 to 2E$_{max}$. This is due to multiparticle interaction effects that are themselves relevant due to long-range dipolar interactions. More importantly, there exist no arguments to expect a transfer of OAM during such a scattering process. 

Finally, we suggest quasi-elastic RS due to down-folding and strong fluctuations from $k_{s}$=$\pi/\textit{l(T)}$. As mentioned before, this signal is related to the locking of thermal to phase fluctuations of the LC that also involve OAM. We assign a Gaussian line-width to this process as it results from a relaxation of non-equilibrium states and decay of topological defects. Such a line-width is typical for solutions of a Fokker Planck equation \cite{Haken}. Microscopic light matter coupling of OAM to such processes are far from trivial as the LG phase front in paraxial regime does not couple to dipolar processes\cite{Araoka,Loeffler1}. However, there exist several loopholes to allow a considerable coupling in higher order, e.g. quadrupolar electric matrix elements, electronic resonances, as well as strong focusing of the LG wavefront\cite{spin-orbit,babiker-02,Forbes-19}. These effects could then be coined as optical spin-orbit coupling \cite{Bliokh}. Please notice that inelastically induced spin-orbit coupling is a well established technique in trapped ion, optical lattice clocks \cite{Li-16,Kolkowitz-17}, and very recently in functionalized Fabry–Perot cavities \cite{Gautier}. While strong focusing can be disregarded for our experiment, higher than dipolar order matrix elements and electronic resonances are definitely relevant for OAM Raman scattering \cite{Forbes-19b}. Concerning topological defects our data is still rather unspecific. So we refrain here from further discussions \cite{Porenta-14,Dierking-14,Tai-19,Papic-21}. 

It is noteworthy that spectral intensities $I_a$$/$$I_b$ presented in Fig. \ref{fig:Fig3} (a) always exhibit positive effects for different OAM/SAM and scattering vectors. This is completely different from ROA \cite{Fischer-05} where phonon maxima of both signs are observed due to interference effects. Hence, we attribute the former case to a selection rule that allows transitions from the left-handed, chiral ground state of the LC to states with an altered angular momentum. 

Summarizing our results, we discriminate helical versus the more common chiral components of a chiral light scattering process. For this instance we used vortex photon fields with a helical phase front and different topological charge that are scattered off chiral LC. The components show up as a finite-energy, more conventional signal as well as a quasi-elastic one, with a strong helical contribution. We emphasize that the polarization of the latter points to a transfer of OAM during the scattering process. Both signals are limited to the parameter range of iridescence within the chiral phase, allowing a certain locking of the scattering vector and the chiral pitch length. The locking leads to a resonant-like optical spin orbit coupling. In our approach it seems critical that we used a Raman optical activity like methodology comparing divided spectra of different OAM/SAM. This leads to an enhanced sensitivity and a cancellation of conventional vibrational modes and background signals. In a broader context, such in-depth work using vortex or structured light is of utmost relevance and linked to current advances in applications ranging from nanophotonics, imaging, metrology, to quantum communication.

\section{Acknowledgements}
We acknowledge support by the Deutsche Forschungsgemeinschaft, DFG Grant No. $LE967/16-1$, DFG GrK 1952/2, Metrology for Complex Nanosystems - NanoMet, and  DFG EXC-2123 QuantumFrontiers – 390837967. Important discussions included A. Surzhykov, W. Löffler, W. Brenig, V. Gnezdilov, and Yu. G. Pashkevich.

  	\bibliographystyle{apsrev4-2}
		\bibliography{biblio}	

\begin{thebibliography}{36}%
\makeatletter
\providecommand \@ifxundefined [1]{%
 \@ifx{#1\undefined}
}%
\providecommand \@ifnum [1]{%
 \ifnum #1\expandafter \@firstoftwo
 \else \expandafter \@secondoftwo
 \fi
}%
\providecommand \@ifx [1]{%
 \ifx #1\expandafter \@firstoftwo
 \else \expandafter \@secondoftwo
 \fi
}%
\providecommand \natexlab [1]{#1}%
\providecommand \enquote  [1]{``#1''}%
\providecommand \bibnamefont  [1]{#1}%
\providecommand \bibfnamefont [1]{#1}%
\providecommand \citenamefont [1]{#1}%
\providecommand \href@noop [0]{\@secondoftwo}%
\providecommand \href [0]{\begingroup \@sanitize@url \@href}%
\providecommand \@href[1]{\@@startlink{#1}\@@href}%
\providecommand \@@href[1]{\endgroup#1\@@endlink}%
\providecommand \@sanitize@url [0]{\catcode `\\12\catcode `\$12\catcode
  `\&12\catcode `\#12\catcode `\^12\catcode `\_12\catcode `\%12\relax}%
\providecommand \@@startlink[1]{}%
\providecommand \@@endlink[0]{}%
\providecommand \url  [0]{\begingroup\@sanitize@url \@url }%
\providecommand \@url [1]{\endgroup\@href {#1}{\urlprefix }}%
\providecommand \urlprefix  [0]{URL }%
\providecommand \Eprint [0]{\href }%
\providecommand \doibase [0]{https://doi.org/}%
\providecommand \selectlanguage [0]{\@gobble}%
\providecommand \bibinfo  [0]{\@secondoftwo}%
\providecommand \bibfield  [0]{\@secondoftwo}%
\providecommand \translation [1]{[#1]}%
\providecommand \BibitemOpen [0]{}%
\providecommand \bibitemStop [0]{}%
\providecommand \bibitemNoStop [0]{.\EOS\space}%
\providecommand \EOS [0]{\spacefactor3000\relax}%
\providecommand \BibitemShut  [1]{\csname bibitem#1\endcsname}%
\let\auto@bib@innerbib\@empty
\bibitem [{\citenamefont {Woods}(1965)}]{NeutronscatterinHe}%
  \BibitemOpen
  \bibfield  {author} {\bibinfo {author} {\bibfnamefont {A.~D.~B.}\
  \bibnamefont {Woods}},\ }\href {https://doi.org/10.1103/PhysRevLett.14.355}
  {\bibfield  {journal} {\bibinfo  {journal} {Phys. Rev. Lett.}\ }\textbf
  {\bibinfo {volume} {14}},\ \bibinfo {pages} {355} (\bibinfo {year}
  {1965})}\BibitemShut {NoStop}%
\bibitem [{\citenamefont {Chomaz}\ \emph {et~al.}(2018)\citenamefont {Chomaz},
  \citenamefont {van Bijnen}, \citenamefont {Petter}, \citenamefont {Faraoni},
  \citenamefont {Baier}, \citenamefont {Becher}, \citenamefont {Mark},
  \citenamefont {Waechtler}, \citenamefont {Santos},\ and\ \citenamefont
  {Ferlaino}}]{dipolargas}%
  \BibitemOpen
  \bibfield  {author} {\bibinfo {author} {\bibfnamefont {L.}~\bibnamefont
  {Chomaz}}, \bibinfo {author} {\bibfnamefont {R.~M.~W.}\ \bibnamefont {van
  Bijnen}}, \bibinfo {author} {\bibfnamefont {D.}~\bibnamefont {Petter}},
  \bibinfo {author} {\bibfnamefont {G.}~\bibnamefont {Faraoni}}, \bibinfo
  {author} {\bibfnamefont {S.}~\bibnamefont {Baier}}, \bibinfo {author}
  {\bibfnamefont {J.~H.}\ \bibnamefont {Becher}}, \bibinfo {author}
  {\bibfnamefont {M.~J.}\ \bibnamefont {Mark}}, \bibinfo {author}
  {\bibfnamefont {F.}~\bibnamefont {Waechtler}}, \bibinfo {author}
  {\bibfnamefont {L.}~\bibnamefont {Santos}},\ and\ \bibinfo {author}
  {\bibfnamefont {F.}~\bibnamefont {Ferlaino}},\ }\href
  {https://doi.org/10.1038/s41567-018-0054-7} {\bibfield  {journal} {\bibinfo
  {journal} {Nature Physics}\ }\textbf {\bibinfo {volume} {14}},\ \bibinfo
  {pages} {442} (\bibinfo {year} {2018})}\BibitemShut {NoStop}%
\bibitem [{\citenamefont {Mottl}\ \emph {et~al.}(2012)\citenamefont {Mottl},
  \citenamefont {Brennecke}, \citenamefont {Baumann}, \citenamefont {Landig},
  \citenamefont {Donner},\ and\ \citenamefont {Esslinger}}]{Rotonsoftening}%
  \BibitemOpen
  \bibfield  {author} {\bibinfo {author} {\bibfnamefont {R.}~\bibnamefont
  {Mottl}}, \bibinfo {author} {\bibfnamefont {F.}~\bibnamefont {Brennecke}},
  \bibinfo {author} {\bibfnamefont {K.}~\bibnamefont {Baumann}}, \bibinfo
  {author} {\bibfnamefont {R.}~\bibnamefont {Landig}}, \bibinfo {author}
  {\bibfnamefont {T.}~\bibnamefont {Donner}},\ and\ \bibinfo {author}
  {\bibfnamefont {T.}~\bibnamefont {Esslinger}},\ }\href
  {https://doi.org/10.1126/science.1220314} {\bibfield  {journal} {\bibinfo
  {journal} {Science}\ }\textbf {\bibinfo {volume} {336}},\ \bibinfo {pages}
  {1570} (\bibinfo {year} {2012})}\BibitemShut {NoStop}%
\bibitem [{\citenamefont {Chen}\ \emph {et~al.}(2021)\citenamefont {Chen},
  \citenamefont {Kadic},\ and\ \citenamefont {Wegener}}]{metamaterial}%
  \BibitemOpen
  \bibfield  {author} {\bibinfo {author} {\bibfnamefont {Y.}~\bibnamefont
  {Chen}}, \bibinfo {author} {\bibfnamefont {M.}~\bibnamefont {Kadic}},\ and\
  \bibinfo {author} {\bibfnamefont {M.}~\bibnamefont {Wegener}},\ }\bibfield
  {journal} {\bibinfo  {journal} {Nature Commun.}\ }\textbf {\bibinfo {volume}
  {12}},\ \href {https://doi.org/10.1038/s41467-021-23574-2}
  {10.1038/s41467-021-23574-2} (\bibinfo {year} {2021})\BibitemShut {NoStop}%
\bibitem [{\citenamefont {Liu}\ \emph {et~al.}(2021)\citenamefont {Liu},
  \citenamefont {Xiao}, \citenamefont {Koo},\ and\ \citenamefont
  {Yan}}]{Liu-21}%
  \BibitemOpen
  \bibfield  {author} {\bibinfo {author} {\bibfnamefont {Y.}~\bibnamefont
  {Liu}}, \bibinfo {author} {\bibfnamefont {J.}~\bibnamefont {Xiao}}, \bibinfo
  {author} {\bibfnamefont {J.}~\bibnamefont {Koo}},\ and\ \bibinfo {author}
  {\bibfnamefont {B.}~\bibnamefont {Yan}},\ }\href
  {https://doi.org/10.1038/s41563-021-00924-5} {\bibfield  {journal} {\bibinfo
  {journal} {Nature Materials}\ }\textbf {\bibinfo {volume} {20}},\ \bibinfo
  {pages} {638} (\bibinfo {year} {2021})}\BibitemShut {NoStop}%
\bibitem [{\citenamefont {Kishine}\ \emph {et~al.}(2020)\citenamefont
  {Kishine}, \citenamefont {Ovchinnikov},\ and\ \citenamefont
  {Tereshchenko}}]{Kishine-20}%
  \BibitemOpen
  \bibfield  {author} {\bibinfo {author} {\bibfnamefont {J.}~\bibnamefont
  {Kishine}}, \bibinfo {author} {\bibfnamefont {A.~S.}\ \bibnamefont
  {Ovchinnikov}},\ and\ \bibinfo {author} {\bibfnamefont {A.~A.}\ \bibnamefont
  {Tereshchenko}},\ }\href {https://doi.org/10.1103/PhysRevLett.125.245302}
  {\bibfield  {journal} {\bibinfo  {journal} {Phys. Rev. Lett.}\ }\textbf
  {\bibinfo {volume} {125}},\ \bibinfo {pages} {245302} (\bibinfo {year}
  {2020})}\BibitemShut {NoStop}%
\bibitem [{\citenamefont {Baghdasaryan}\ \emph {et~al.}(2022)\citenamefont
  {Baghdasaryan}, \citenamefont {Hakobyan}, \citenamefont {Hayrapetyan},
  \citenamefont {Iorsh}, \citenamefont {Shelykh},\ and\ \citenamefont
  {Shahnazaryan}}]{Baghda-22}%
  \BibitemOpen
  \bibfield  {author} {\bibinfo {author} {\bibfnamefont {D.~A.}\ \bibnamefont
  {Baghdasaryan}}, \bibinfo {author} {\bibfnamefont {E.~S.}\ \bibnamefont
  {Hakobyan}}, \bibinfo {author} {\bibfnamefont {D.~B.}\ \bibnamefont
  {Hayrapetyan}}, \bibinfo {author} {\bibfnamefont {I.~V.}\ \bibnamefont
  {Iorsh}}, \bibinfo {author} {\bibfnamefont {I.~A.}\ \bibnamefont {Shelykh}},\
  and\ \bibinfo {author} {\bibfnamefont {V.}~\bibnamefont {Shahnazaryan}},\
  }\href {https://doi.org/10.1103/PhysRevMaterials.6.034003} {\bibfield
  {journal} {\bibinfo  {journal} {Phys. Rev. Materials}\ }\textbf {\bibinfo
  {volume} {6}},\ \bibinfo {pages} {034003} (\bibinfo {year}
  {2022})}\BibitemShut {NoStop}%
\bibitem [{\citenamefont {Rubinsztein-Dunlop}\ \emph
  {et~al.}(2016)\citenamefont {Rubinsztein-Dunlop}, \citenamefont {Forbes},
  \citenamefont {Berry}, \citenamefont {Dennis}, \citenamefont {Andrews},
  \citenamefont {Mansuripur}, \citenamefont {Denz}, \citenamefont {Alpmann},
  \citenamefont {Banzer}, \citenamefont {Bauer}, \citenamefont {Karimi},
  \citenamefont {Marrucci}, \citenamefont {Padgett}, \citenamefont
  {Ritsch-Marte}, \citenamefont {Litchinitser}, \citenamefont {Bigelow},
  \citenamefont {Rosales-Guzm{\'{a}}n}, \citenamefont {Belmonte}, \citenamefont
  {Torres}, \citenamefont {Neely}, \citenamefont {Baker}, \citenamefont
  {Gordon}, \citenamefont {Stilgoe}, \citenamefont {Romero}, \citenamefont
  {White}, \citenamefont {Fickler}, \citenamefont {Willner}, \citenamefont
  {Xie}, \citenamefont {McMorran},\ and\ \citenamefont {Weiner}}]{Rubinsztein}%
  \BibitemOpen
  \bibfield  {author} {\bibinfo {author} {\bibfnamefont {H.}~\bibnamefont
  {Rubinsztein-Dunlop}}, \bibinfo {author} {\bibfnamefont {A.}~\bibnamefont
  {Forbes}}, \bibinfo {author} {\bibfnamefont {M.~V.}\ \bibnamefont {Berry}},
  \bibinfo {author} {\bibfnamefont {M.~R.}\ \bibnamefont {Dennis}}, \bibinfo
  {author} {\bibfnamefont {D.~L.}\ \bibnamefont {Andrews}}, \bibinfo {author}
  {\bibfnamefont {M.}~\bibnamefont {Mansuripur}}, \bibinfo {author}
  {\bibfnamefont {C.}~\bibnamefont {Denz}}, \bibinfo {author} {\bibfnamefont
  {C.}~\bibnamefont {Alpmann}}, \bibinfo {author} {\bibfnamefont
  {P.}~\bibnamefont {Banzer}}, \bibinfo {author} {\bibfnamefont
  {T.}~\bibnamefont {Bauer}}, \bibinfo {author} {\bibfnamefont
  {E.}~\bibnamefont {Karimi}}, \bibinfo {author} {\bibfnamefont
  {L.}~\bibnamefont {Marrucci}}, \bibinfo {author} {\bibfnamefont
  {M.}~\bibnamefont {Padgett}}, \bibinfo {author} {\bibfnamefont
  {M.}~\bibnamefont {Ritsch-Marte}}, \bibinfo {author} {\bibfnamefont {N.~M.}\
  \bibnamefont {Litchinitser}}, \bibinfo {author} {\bibfnamefont {N.~P.}\
  \bibnamefont {Bigelow}}, \bibinfo {author} {\bibfnamefont {C.}~\bibnamefont
  {Rosales-Guzm{\'{a}}n}}, \bibinfo {author} {\bibfnamefont {A.}~\bibnamefont
  {Belmonte}}, \bibinfo {author} {\bibfnamefont {J.~P.}\ \bibnamefont
  {Torres}}, \bibinfo {author} {\bibfnamefont {T.~W.}\ \bibnamefont {Neely}},
  \bibinfo {author} {\bibfnamefont {M.}~\bibnamefont {Baker}}, \bibinfo
  {author} {\bibfnamefont {R.}~\bibnamefont {Gordon}}, \bibinfo {author}
  {\bibfnamefont {A.~B.}\ \bibnamefont {Stilgoe}}, \bibinfo {author}
  {\bibfnamefont {J.}~\bibnamefont {Romero}}, \bibinfo {author} {\bibfnamefont
  {A.~G.}\ \bibnamefont {White}}, \bibinfo {author} {\bibfnamefont
  {R.}~\bibnamefont {Fickler}}, \bibinfo {author} {\bibfnamefont {A.~E.}\
  \bibnamefont {Willner}}, \bibinfo {author} {\bibfnamefont {G.}~\bibnamefont
  {Xie}}, \bibinfo {author} {\bibfnamefont {B.}~\bibnamefont {McMorran}},\ and\
  \bibinfo {author} {\bibfnamefont {A.~M.}\ \bibnamefont {Weiner}},\ }\href
  {https://doi.org/10.1088/2040-8978/19/1/013001} {\bibfield  {journal}
  {\bibinfo  {journal} {J. of Optics}\ }\textbf {\bibinfo {volume} {19}},\
  \bibinfo {pages} {013001} (\bibinfo {year} {2016})}\BibitemShut {NoStop}%
\bibitem [{\citenamefont {Padgett}\ and\ \citenamefont
  {Bowman}(2011)}]{tweezers}%
  \BibitemOpen
  \bibfield  {author} {\bibinfo {author} {\bibfnamefont {M.~J.}\ \bibnamefont
  {Padgett}}\ and\ \bibinfo {author} {\bibfnamefont {R.~W.}\ \bibnamefont
  {Bowman}},\ }\href@noop {} {\bibfield  {journal} {\bibinfo  {journal} {Nature
  Photonics}\ }\textbf {\bibinfo {volume} {5}},\ \bibinfo {pages} {343}
  (\bibinfo {year} {2011})}\BibitemShut {NoStop}%
\bibitem [{\citenamefont {Tang}\ and\ \citenamefont {Cohen}(2010)}]{enhanced1}%
  \BibitemOpen
  \bibfield  {author} {\bibinfo {author} {\bibfnamefont {Y.}~\bibnamefont
  {Tang}}\ and\ \bibinfo {author} {\bibfnamefont {A.~E.}\ \bibnamefont
  {Cohen}},\ }\href {https://doi.org/10.1103/PhysRevLett.104.163901} {\bibfield
   {journal} {\bibinfo  {journal} {Phys. Rev. Lett.}\ }\textbf {\bibinfo
  {volume} {104}},\ \bibinfo {pages} {163901} (\bibinfo {year}
  {2010})}\BibitemShut {NoStop}%
\bibitem [{\citenamefont {Rosales-Guzm{\'{a}}n}\ \emph
  {et~al.}(2012)\citenamefont {Rosales-Guzm{\'{a}}n}, \citenamefont
  {Volke-Sepulveda},\ and\ \citenamefont {Torres}}]{enhanced2}%
  \BibitemOpen
  \bibfield  {author} {\bibinfo {author} {\bibfnamefont {C.}~\bibnamefont
  {Rosales-Guzm{\'{a}}n}}, \bibinfo {author} {\bibfnamefont {K.}~\bibnamefont
  {Volke-Sepulveda}},\ and\ \bibinfo {author} {\bibfnamefont {J.~P.}\
  \bibnamefont {Torres}},\ }\href {https://doi.org/10.1364/OL.37.003486}
  {\bibfield  {journal} {\bibinfo  {journal} {Opt Lett.}\ }\textbf {\bibinfo
  {volume} {37}},\ \bibinfo {pages} {3486} (\bibinfo {year}
  {2012})}\BibitemShut {NoStop}%
\bibitem [{\citenamefont {Forbes}\ and\ \citenamefont
  {Andrews}(2019)}]{enhanced3}%
  \BibitemOpen
  \bibfield  {author} {\bibinfo {author} {\bibfnamefont {K.~A.}\ \bibnamefont
  {Forbes}}\ and\ \bibinfo {author} {\bibfnamefont {D.~L.}\ \bibnamefont
  {Andrews}},\ }\href {https://doi.org/10.1103/PhysRevResearch.1.033080}
  {\bibfield  {journal} {\bibinfo  {journal} {Phys. Rev. Research}\ }\textbf
  {\bibinfo {volume} {1}},\ \bibinfo {pages} {033080} (\bibinfo {year}
  {2019})}\BibitemShut {NoStop}%
\bibitem [{\citenamefont {L\"offler}\ \emph {et~al.}(2011)\citenamefont
  {L\"offler}, \citenamefont {Broer},\ and\ \citenamefont
  {Woerdman}}]{Loeffler1}%
  \BibitemOpen
  \bibfield  {author} {\bibinfo {author} {\bibfnamefont {W.}~\bibnamefont
  {L\"offler}}, \bibinfo {author} {\bibfnamefont {D.~J.}\ \bibnamefont
  {Broer}},\ and\ \bibinfo {author} {\bibfnamefont {J.~P.}\ \bibnamefont
  {Woerdman}},\ }\href {https://doi.org/10.1103/PhysRevA.83.065801} {\bibfield
  {journal} {\bibinfo  {journal} {Phys. Rev. A}\ }\textbf {\bibinfo {volume}
  {83}},\ \bibinfo {pages} {065801} (\bibinfo {year} {2011})}\BibitemShut
  {NoStop}%
\bibitem [{\citenamefont {Araoka}\ \emph {et~al.}(2005)\citenamefont {Araoka},
  \citenamefont {Verbiest}, \citenamefont {Clays},\ and\ \citenamefont
  {Persoons}}]{Araoka}%
  \BibitemOpen
  \bibfield  {author} {\bibinfo {author} {\bibfnamefont {F.}~\bibnamefont
  {Araoka}}, \bibinfo {author} {\bibfnamefont {T.}~\bibnamefont {Verbiest}},
  \bibinfo {author} {\bibfnamefont {K.}~\bibnamefont {Clays}},\ and\ \bibinfo
  {author} {\bibfnamefont {A.}~\bibnamefont {Persoons}},\ }\href
  {https://doi.org/10.1103/PhysRevA.71.055401} {\bibfield  {journal} {\bibinfo
  {journal} {Phys. Rev. A}\ }\textbf {\bibinfo {volume} {71}},\ \bibinfo
  {pages} {055401} (\bibinfo {year} {2005})}\BibitemShut {NoStop}%
\bibitem [{\citenamefont {Woźniak}\ \emph {et~al.}(2019)\citenamefont
  {Woźniak}, \citenamefont {De~Leon}, \citenamefont {Höflich}, \citenamefont
  {Leuchs},\ and\ \citenamefont {Banzer}}]{Wozniak}%
  \BibitemOpen
  \bibfield  {author} {\bibinfo {author} {\bibfnamefont {P.}~\bibnamefont
  {Woźniak}}, \bibinfo {author} {\bibfnamefont {I.}~\bibnamefont {De~Leon}},
  \bibinfo {author} {\bibfnamefont {K.}~\bibnamefont {Höflich}}, \bibinfo
  {author} {\bibfnamefont {G.}~\bibnamefont {Leuchs}},\ and\ \bibinfo {author}
  {\bibfnamefont {P.}~\bibnamefont {Banzer}},\ }\href
  {https://doi.org/10.1364/OPTICA.6.000961} {\bibfield  {journal} {\bibinfo
  {journal} {Optica}\ }\textbf {\bibinfo {volume} {6}},\ \bibinfo {pages} {961}
  (\bibinfo {year} {2019})}\BibitemShut {NoStop}%
\bibitem [{\citenamefont {Forbes}\ and\ \citenamefont
  {Andrews}(2021)}]{Forbes-21b}%
  \BibitemOpen
  \bibfield  {author} {\bibinfo {author} {\bibfnamefont {K.~A.}\ \bibnamefont
  {Forbes}}\ and\ \bibinfo {author} {\bibfnamefont {D.~L.}\ \bibnamefont
  {Andrews}},\ }\href {https://doi.org/10.1088/2515-7647/abdb06} {\bibfield
  {journal} {\bibinfo  {journal} {J. of Physics: Photonics}\ }\textbf {\bibinfo
  {volume} {3}},\ \bibinfo {pages} {022007} (\bibinfo {year}
  {2021})}\BibitemShut {NoStop}%
\bibitem [{\citenamefont {Bliokh}\ \emph {et~al.}(2010)\citenamefont {Bliokh},
  \citenamefont {Alonso}, \citenamefont {Ostrovskaya},\ and\ \citenamefont
  {Aiello}}]{spin-orbit}%
  \BibitemOpen
  \bibfield  {author} {\bibinfo {author} {\bibfnamefont {K.~Y.}\ \bibnamefont
  {Bliokh}}, \bibinfo {author} {\bibfnamefont {M.~A.}\ \bibnamefont {Alonso}},
  \bibinfo {author} {\bibfnamefont {E.~A.}\ \bibnamefont {Ostrovskaya}},\ and\
  \bibinfo {author} {\bibfnamefont {A.}~\bibnamefont {Aiello}},\ }\href
  {https://doi.org/10.1103/PhysRevA.82.063825} {\bibfield  {journal} {\bibinfo
  {journal} {Phys. Rev. A}\ }\textbf {\bibinfo {volume} {82}},\ \bibinfo
  {pages} {063825} (\bibinfo {year} {2010})}\BibitemShut {NoStop}%
\bibitem [{\citenamefont {Forbes}\ and\ \citenamefont
  {Jones}(2021)}]{Forbes-21}%
  \BibitemOpen
  \bibfield  {author} {\bibinfo {author} {\bibfnamefont {K.~A.}\ \bibnamefont
  {Forbes}}\ and\ \bibinfo {author} {\bibfnamefont {G.~A.}\ \bibnamefont
  {Jones}},\ }\href {https://doi.org/10.1103/PhysRevA.103.053515} {\bibfield
  {journal} {\bibinfo  {journal} {Phys. Rev. A}\ }\textbf {\bibinfo {volume}
  {103}},\ \bibinfo {pages} {053515} (\bibinfo {year} {2021})}\BibitemShut
  {NoStop}%
\bibitem [{\citenamefont {Khoo}(2009)}]{Khoo-09}%
  \BibitemOpen
  \bibfield  {author} {\bibinfo {author} {\bibfnamefont {I.~C.}\ \bibnamefont
  {Khoo}},\ }\href
  {https://doi.org/https://doi.org/10.1016/j.physrep.2009.01.001} {\bibfield
  {journal} {\bibinfo  {journal} {Physics Reports}\ }\textbf {\bibinfo {volume}
  {471}},\ \bibinfo {pages} {221} (\bibinfo {year} {2009})}\BibitemShut
  {NoStop}%
\bibitem [{\citenamefont {Zhang}\ \emph {et~al.}(2021)\citenamefont {Zhang},
  \citenamefont {Froyen}, \citenamefont {Schenning}, \citenamefont {Zhou},
  \citenamefont {Debije},\ and\ \citenamefont {de~Haan}}]{Zhang-21}%
  \BibitemOpen
  \bibfield  {author} {\bibinfo {author} {\bibfnamefont {W.}~\bibnamefont
  {Zhang}}, \bibinfo {author} {\bibfnamefont {A.~A.~F.}\ \bibnamefont
  {Froyen}}, \bibinfo {author} {\bibfnamefont {A.~P. H.~J.}\ \bibnamefont
  {Schenning}}, \bibinfo {author} {\bibfnamefont {G.}~\bibnamefont {Zhou}},
  \bibinfo {author} {\bibfnamefont {M.~G.}\ \bibnamefont {Debije}},\ and\
  \bibinfo {author} {\bibfnamefont {L.~T.}\ \bibnamefont {de~Haan}},\ }\href
  {https://doi.org/doi.org/10.1002/adpr.202100016} {\bibfield  {journal}
  {\bibinfo  {journal} {Adv. Photonics Res.}\ }\textbf {\bibinfo {volume}
  {2}},\ \bibinfo {pages} {2100016} (\bibinfo {year} {2021})}\BibitemShut
  {NoStop}%
\bibitem [{\citenamefont {Papič}\ \emph {et~al.}(2021)\citenamefont {Papič},
  \citenamefont {Mur}, \citenamefont {Zuhail}, \citenamefont {Ravnik},
  \citenamefont {Muševič},\ and\ \citenamefont {Humar}}]{Papic-21}%
  \BibitemOpen
  \bibfield  {author} {\bibinfo {author} {\bibfnamefont {M.}~\bibnamefont
  {Papič}}, \bibinfo {author} {\bibfnamefont {U.}~\bibnamefont {Mur}},
  \bibinfo {author} {\bibfnamefont {K.~P.}\ \bibnamefont {Zuhail}}, \bibinfo
  {author} {\bibfnamefont {M.}~\bibnamefont {Ravnik}}, \bibinfo {author}
  {\bibfnamefont {I.}~\bibnamefont {Muševič}},\ and\ \bibinfo {author}
  {\bibfnamefont {M.}~\bibnamefont {Humar}},\ }\href
  {https://doi.org/10.1073/pnas.2110839118} {\bibfield  {journal} {\bibinfo
  {journal} {Proceedings of the National Academy of Sciences}\ }\textbf
  {\bibinfo {volume} {118}},\ \bibinfo {pages} {e2110839118} (\bibinfo {year}
  {2021})},\ \Eprint
  {https://arxiv.org/abs/https://www.pnas.org/doi/pdf/10.1073/pnas.2110839118}
  {https://www.pnas.org/doi/pdf/10.1073/pnas.2110839118} \BibitemShut {NoStop}%
\bibitem [{\citenamefont {Forbes}(2019)}]{Forbes-19}%
  \BibitemOpen
  \bibfield  {author} {\bibinfo {author} {\bibfnamefont {K.~A.}\ \bibnamefont
  {Forbes}},\ }\href {https://doi.org/10.1103/PhysRevLett.122.103201}
  {\bibfield  {journal} {\bibinfo  {journal} {Phys. Rev. Lett.}\ }\textbf
  {\bibinfo {volume} {122}},\ \bibinfo {pages} {103201} (\bibinfo {year}
  {2019})}\BibitemShut {NoStop}%
\bibitem [{\citenamefont {Büscher}\ \emph {et~al.}(2021)\citenamefont
  {Büscher}, \citenamefont {Müllner}, \citenamefont {Wulferding},
  \citenamefont {Pashkevich}, \citenamefont {Gnezdilov}, \citenamefont
  {Peshkov}, \citenamefont {Surzhykov},\ and\ \citenamefont
  {Lemmens}}]{Florian}%
  \BibitemOpen
  \bibfield  {author} {\bibinfo {author} {\bibfnamefont {F.}~\bibnamefont
  {Büscher}}, \bibinfo {author} {\bibfnamefont {S.}~\bibnamefont {Müllner}},
  \bibinfo {author} {\bibfnamefont {D.}~\bibnamefont {Wulferding}}, \bibinfo
  {author} {\bibfnamefont {Y.~G.}\ \bibnamefont {Pashkevich}}, \bibinfo
  {author} {\bibfnamefont {V.}~\bibnamefont {Gnezdilov}}, \bibinfo {author}
  {\bibfnamefont {A.~A.}\ \bibnamefont {Peshkov}}, \bibinfo {author}
  {\bibfnamefont {A.}~\bibnamefont {Surzhykov}},\ and\ \bibinfo {author}
  {\bibfnamefont {P.}~\bibnamefont {Lemmens}},\ }\href
  {https://doi.org/10.1063/10.0006577} {\bibfield  {journal} {\bibinfo
  {journal} {Low Temp. Phys.}\ }\textbf {\bibinfo {volume} {47}},\ \bibinfo
  {pages} {959} (\bibinfo {year} {2021})}\BibitemShut {NoStop}%
\bibitem [{\citenamefont {Nozi{\'{e}}re}(2004)}]{Noziere-04}%
  \BibitemOpen
  \bibfield  {author} {\bibinfo {author} {\bibfnamefont {P.}~\bibnamefont
  {Nozi{\'{e}}re}},\ }\href
  {https://doi.org/10.1023/B:JOLT.0000044234.82957.2f} {\bibfield  {journal}
  {\bibinfo  {journal} {J. Low. Temp. Phys.}\ }\textbf {\bibinfo {volume}
  {137}},\ \bibinfo {pages} {45} (\bibinfo {year} {2004})}\BibitemShut
  {NoStop}%
\bibitem [{\citenamefont {Nozi{\'{e}}re}(2009)}]{Noziere-09}%
  \BibitemOpen
  \bibfield  {author} {\bibinfo {author} {\bibfnamefont {P.}~\bibnamefont
  {Nozi{\'{e}}re}},\ }\href {https://doi.org/10.1007/s10909-009-9889-8}
  {\bibfield  {journal} {\bibinfo  {journal} {J. Low. Temp. Phys.}\ }\textbf
  {\bibinfo {volume} {156}},\ \bibinfo {pages} {9} (\bibinfo {year}
  {2009})}\BibitemShut {NoStop}%
\bibitem [{\citenamefont {Haken}(1977)}]{Haken}%
  \BibitemOpen
  \bibfield  {author} {\bibinfo {author} {\bibfnamefont {H.}~\bibnamefont
  {Haken}},\ }\href@noop {} {\emph {\bibinfo {title} {Synergetics. An
  Introduction}}}\ (\bibinfo  {publisher} {Springer-Verlag},\ \bibinfo
  {address} {Berlin-Heidelberg-New York},\ \bibinfo {year} {1977})\BibitemShut
  {NoStop}%
\bibitem [{\citenamefont {Babiker}\ \emph {et~al.}(2002)\citenamefont
  {Babiker}, \citenamefont {Bennett}, \citenamefont {Andrews},\ and\
  \citenamefont {D\'avila~Romero}}]{babiker-02}%
  \BibitemOpen
  \bibfield  {author} {\bibinfo {author} {\bibfnamefont {M.}~\bibnamefont
  {Babiker}}, \bibinfo {author} {\bibfnamefont {C.~R.}\ \bibnamefont
  {Bennett}}, \bibinfo {author} {\bibfnamefont {D.~L.}\ \bibnamefont
  {Andrews}},\ and\ \bibinfo {author} {\bibfnamefont {L.~C.}\ \bibnamefont
  {D\'avila~Romero}},\ }\href {https://doi.org/10.1103/PhysRevLett.89.143601}
  {\bibfield  {journal} {\bibinfo  {journal} {Phys. Rev. Lett.}\ }\textbf
  {\bibinfo {volume} {89}},\ \bibinfo {pages} {143601} (\bibinfo {year}
  {2002})}\BibitemShut {NoStop}%
\bibitem [{\citenamefont {Bliokh}\ \emph {et~al.}(2015)\citenamefont {Bliokh},
  \citenamefont {Rodríguez-Fortu\~{n}o}, \citenamefont {Nori},\ and\
  \citenamefont {Zayats}}]{Bliokh}%
  \BibitemOpen
  \bibfield  {author} {\bibinfo {author} {\bibfnamefont {K.~Y.}\ \bibnamefont
  {Bliokh}}, \bibinfo {author} {\bibfnamefont {F.~J.}\ \bibnamefont
  {Rodríguez-Fortu\~{n}o}}, \bibinfo {author} {\bibfnamefont {F.}~\bibnamefont
  {Nori}},\ and\ \bibinfo {author} {\bibfnamefont {A.~V.}\ \bibnamefont
  {Zayats}},\ }\href {https://doi.org/10.1038/nphoton.2015.201} {\bibfield
  {journal} {\bibinfo  {journal} {Nature Photonics}\ }\textbf {\bibinfo
  {volume} {9}},\ \bibinfo {pages} {796} (\bibinfo {year} {2015})}\BibitemShut
  {NoStop}%
\bibitem [{\citenamefont {Li}\ \emph {et~al.}(2016)\citenamefont {Li},
  \citenamefont {Huang}, \citenamefont {Shteynas}, \citenamefont {Burchesky},
  \citenamefont {Top}, \citenamefont {Su}, \citenamefont {Lee}, \citenamefont
  {Jamison},\ and\ \citenamefont {Ketterle}}]{Li-16}%
  \BibitemOpen
  \bibfield  {author} {\bibinfo {author} {\bibfnamefont {J.}~\bibnamefont
  {Li}}, \bibinfo {author} {\bibfnamefont {W.}~\bibnamefont {Huang}}, \bibinfo
  {author} {\bibfnamefont {B.}~\bibnamefont {Shteynas}}, \bibinfo {author}
  {\bibfnamefont {S.}~\bibnamefont {Burchesky}}, \bibinfo {author}
  {\bibfnamefont {F.~C.}\ \bibnamefont {Top}}, \bibinfo {author} {\bibfnamefont
  {E.}~\bibnamefont {Su}}, \bibinfo {author} {\bibfnamefont {J.}~\bibnamefont
  {Lee}}, \bibinfo {author} {\bibfnamefont {A.~O.}\ \bibnamefont {Jamison}},\
  and\ \bibinfo {author} {\bibfnamefont {W.}~\bibnamefont {Ketterle}},\ }\href
  {https://doi.org/10.1103/PhysRevLett.117.185301} {\bibfield  {journal}
  {\bibinfo  {journal} {Phys. Rev. Lett.}\ }\textbf {\bibinfo {volume} {117}},\
  \bibinfo {pages} {185301} (\bibinfo {year} {2016})}\BibitemShut {NoStop}%
\bibitem [{\citenamefont {Kolkowitz}\ \emph {et~al.}(2017)\citenamefont
  {Kolkowitz}, \citenamefont {Bromley}, \citenamefont {Bothwell}, \citenamefont
  {Wall}, \citenamefont {Marti}, \citenamefont {Koller}, \citenamefont {Zhang},
  \citenamefont {M.},\ and\ \citenamefont {J.}}]{Kolkowitz-17}%
  \BibitemOpen
  \bibfield  {author} {\bibinfo {author} {\bibfnamefont {S.}~\bibnamefont
  {Kolkowitz}}, \bibinfo {author} {\bibfnamefont {S.~L.}\ \bibnamefont
  {Bromley}}, \bibinfo {author} {\bibfnamefont {T.}~\bibnamefont {Bothwell}},
  \bibinfo {author} {\bibfnamefont {M.~L.}\ \bibnamefont {Wall}}, \bibinfo
  {author} {\bibfnamefont {G.~E.}\ \bibnamefont {Marti}}, \bibinfo {author}
  {\bibfnamefont {A.~P.}\ \bibnamefont {Koller}}, \bibinfo {author}
  {\bibfnamefont {X.}~\bibnamefont {Zhang}}, \bibinfo {author} {\bibfnamefont
  {R.~A.}\ \bibnamefont {M.}},\ and\ \bibinfo {author} {\bibfnamefont
  {Y.}~\bibnamefont {J.}},\ }\href@noop {} {\bibfield  {journal} {\bibinfo
  {journal} {Nature}\ }\textbf {\bibinfo {volume} {542}} (\bibinfo {year}
  {2017})}\BibitemShut {NoStop}%
\bibitem [{\citenamefont {Gautier}\ \emph {et~al.}(2022)\citenamefont
  {Gautier}, \citenamefont {Li}, \citenamefont {Ebbesen},\ and\ \citenamefont
  {Genet}}]{Gautier}%
  \BibitemOpen
  \bibfield  {author} {\bibinfo {author} {\bibfnamefont {J.}~\bibnamefont
  {Gautier}}, \bibinfo {author} {\bibfnamefont {M.}~\bibnamefont {Li}},
  \bibinfo {author} {\bibfnamefont {T.~W.}\ \bibnamefont {Ebbesen}},\ and\
  \bibinfo {author} {\bibfnamefont {C.}~\bibnamefont {Genet}},\ }\href
  {https://doi.org/10.1021/acsphotonics.1c00780} {\bibfield  {journal}
  {\bibinfo  {journal} {ACS Photonics}\ }\textbf {\bibinfo {volume} {9}},\
  \bibinfo {pages} {778} (\bibinfo {year} {2022})}\BibitemShut {NoStop}%
\bibitem [{\citenamefont {Forbes}\ and\ \citenamefont
  {Salam}(2019)}]{Forbes-19b}%
  \BibitemOpen
  \bibfield  {author} {\bibinfo {author} {\bibfnamefont {K.~A.}\ \bibnamefont
  {Forbes}}\ and\ \bibinfo {author} {\bibfnamefont {A.}~\bibnamefont {Salam}},\
  }\href {https://doi.org/10.1103/PhysRevA.100.053413} {\bibfield  {journal}
  {\bibinfo  {journal} {Phys. Rev. A}\ }\textbf {\bibinfo {volume} {100}},\
  \bibinfo {pages} {053413} (\bibinfo {year} {2019})}\BibitemShut {NoStop}%
\bibitem [{\citenamefont {Porenta}\ \emph {et~al.}(2014)\citenamefont
  {Porenta}, \citenamefont {Copar}, \citenamefont {Ackerman}, \citenamefont
  {Pandey}, \citenamefont {Varney}, \citenamefont {Smalyukh},\ and\
  \citenamefont {Zumer}}]{Porenta-14}%
  \BibitemOpen
  \bibfield  {author} {\bibinfo {author} {\bibfnamefont {T.}~\bibnamefont
  {Porenta}}, \bibinfo {author} {\bibfnamefont {S.}~\bibnamefont {Copar}},
  \bibinfo {author} {\bibfnamefont {P.~J.}\ \bibnamefont {Ackerman}}, \bibinfo
  {author} {\bibfnamefont {M.~B.}\ \bibnamefont {Pandey}}, \bibinfo {author}
  {\bibfnamefont {M.~C.~M.}\ \bibnamefont {Varney}}, \bibinfo {author}
  {\bibfnamefont {I.~I.}\ \bibnamefont {Smalyukh}},\ and\ \bibinfo {author}
  {\bibfnamefont {S.}~\bibnamefont {Zumer}},\ }\href {https://doi.org/DOI:
  10.1038/srep07337} {\bibfield  {journal} {\bibinfo  {journal} {Scientific
  Reports}\ }\textbf {\bibinfo {volume} {4}},\ \bibinfo {pages} {7337}
  (\bibinfo {year} {2014})}\BibitemShut {NoStop}%
\bibitem [{\citenamefont {Dierking}(2014)}]{Dierking-14}%
  \BibitemOpen
  \bibfield  {author} {\bibinfo {author} {\bibfnamefont {I.}~\bibnamefont
  {Dierking}},\ }\href {https://doi.org/10.1021/acsphotonics.1c00780}
  {\bibfield  {journal} {\bibinfo  {journal} {Symmetry}\ }\textbf {\bibinfo
  {volume} {6}},\ \bibinfo {pages} {444} (\bibinfo {year} {2014})}\BibitemShut
  {NoStop}%
\bibitem [{\citenamefont {Tai}\ and\ \citenamefont {Smalyukh}(2019)}]{Tai-19}%
  \BibitemOpen
  \bibfield  {author} {\bibinfo {author} {\bibfnamefont {J.-S.~B.}\
  \bibnamefont {Tai}}\ and\ \bibinfo {author} {\bibfnamefont {I.~I.}\
  \bibnamefont {Smalyukh}},\ }\href {https://doi.org/DOI:
  10.1126/science.aay1638} {\bibfield  {journal} {\bibinfo  {journal}
  {Science}\ }\textbf {\bibinfo {volume} {365}},\ \bibinfo {pages} {1449}
  (\bibinfo {year} {2019})}\BibitemShut {NoStop}%
\bibitem [{\citenamefont {Fischer}\ and\ \citenamefont
  {Hache}(2005)}]{Fischer-05}%
  \BibitemOpen
  \bibfield  {author} {\bibinfo {author} {\bibfnamefont {P.}~\bibnamefont
  {Fischer}}\ and\ \bibinfo {author} {\bibfnamefont {F.}~\bibnamefont
  {Hache}},\ }\href {https://doi.org/10.1002/chir.20179} {\bibfield  {journal}
  {\bibinfo  {journal} {Chirality}\ }\textbf {\bibinfo {volume} {17}},\
  \bibinfo {pages} {421} (\bibinfo {year} {2005})}\BibitemShut {NoStop}%
\end{thebibliography}%
\end{document}